\documentclass[10pt,preprint,showpacs,showkeys,preprintnumbers,superscriptaddress,amsmath,amssymb,twocolumn,prl]{revtex4}
\usepackage{graphicx}
\usepackage{dcolumn}
\usepackage{bm}
\usepackage{epsfig}
\usepackage{amsmath}
\usepackage{amsfonts}
\usepackage{amssymb}
\usepackage{times}

\begin{document}

\preprint{}
\title{Destabilization of the Mg-H system through elastic constraints}
\author{A.~Baldi}
\email[Corresponding author email: abaldi@few.vu.nl]{}
\author{M.~Gonzalez-Silveira}
\author{V.~Palmisano}
\affiliation{Department of Physics and Astronomy, VU University Amsterdam, De Boelelaan 1081, 1081 HV Amsterdam, The Netherlands}
\author{B.~Dam}
\affiliation{DelftChemTech, Technical University Delft, Julianalaan 136, 2600 GA Delft, The Netherlands}
\author{R.~Griessen}
\affiliation{Department of Physics and Astronomy, VU University Amsterdam, De Boelelaan 1081, 1081 HV Amsterdam, The Netherlands}

\date{\today}
\preprint{}
\begin{abstract}
We tune the thermodynamics of hydrogen absorption in Mg by means of elastic clamping. The loading isotherms measured by hydrogenography
show that Mg films covered with Mg-alloy-forming elements, such as Pd and Ni, have hydrogen plateau pressures more than two orders of
magnitude higher than bulk Mg at the same temperature. An elastic model allows us to interpret the Mg thickness dependence of the hydrogen
plateau pressure. Our results provide the basis for the development of new hydrogen storage materials with excellent thermodynamic
properties.
\end{abstract}



\maketitle

Hydrogen is an attractive energy carrier for a future sustainable energy system. Compact hydrogen storage is however still a scientific and
technological challenge \cite{Schlapbach2001}. Storage of molecular hydrogen, both gaseous and liquid, requires high pressures or very low
temperatures and is thus not energy efficient. An alternative is to store atomic hydrogen in metal or complex hydrides. Storage in
metal-hydrides allows, in principle, to achieve high volumetric and gravimetric densities and to reversibly operate at room temperature and
atmospheric pressure. However there are thermodynamic and kinetic limitations associated with the chemical reactions involved in hydrogen
absorption and desorption.

An optimal metal-hydride system has a hydrogen equilibrium pressure of 1 bar at room temperature. Due to weight constraints the interest of
the scientific community has turned to lightweight hydride-forming elements such as Li, B, Na, Mg and Al. Mg can store up to 7.6 mass\%{}
of hydrogen, forming the ionic MgH$_2$ compound, but shows poor kinetics of hydrogen absorption/desorption and a hydrogen equilibrium
pressure of 1 bar at the far too high temperature of 573 K. The absorption and desorption kinetics can be enhanced by reducing the size of
the Mg grains via ball milling \cite{Zaluska1999} hence shortening the hydrogen diffusion length, or by addition of proper catalysts
\cite{Oelerich2001}. Many efforts are currently dedicated to modifying the thermodynamics of the Mg-H system in order to promote hydrogen
dissociation at lower temperatures. Some destabilization has been observed upon reduction of particle size \cite{VarinJALCOM2006} and
alloying of Mg with other elements \cite{Vajo2004,Gremaud2007ADM}, although this generally implies a reduction in the total hydrogen storage
capacity. Theoretical works suggest the possibility to reduce the hydrogen desorption temperature in both MgH$_2$ nanoparticles with grain
size smaller than $\sim$1.3 nm \cite{Wagemans2005JACS} and in MgH$_2$ layers with thicknesses below few unit cells \cite{Liang2005}.

\begin{figure}[htbp]
\begin{center}
\includegraphics[width=0.45\textwidth]{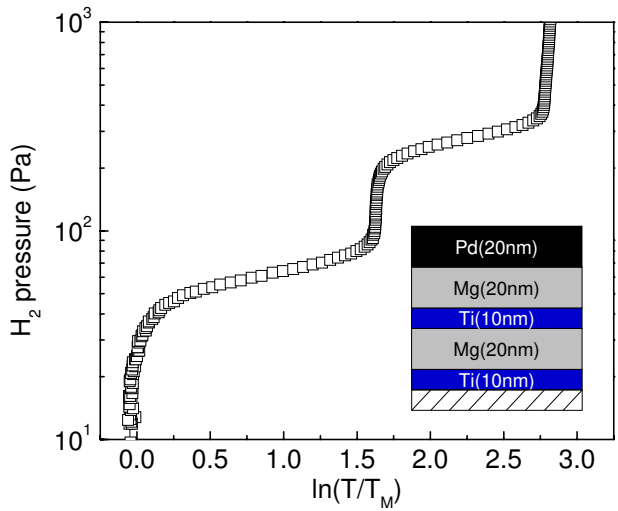}
\caption{(color online) Loading PTI measured at 333 K for a 2x[Ti(10nm)Mg(20nm)]Pd(10nm) sample deposited on glass. In the inset a cartoon
of the sample geometry: glass-substrate/Ti/Mg/Ti/Mg/Pd.}\label{bilayer}
\end{center}
\end{figure}

\begin{figure}[htbp]
\begin{center}
\includegraphics[width=0.45\textwidth]{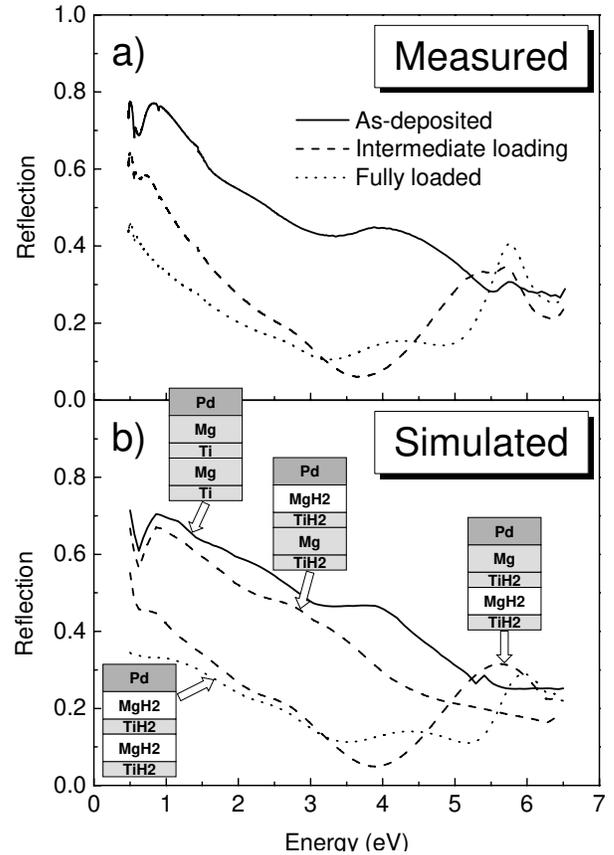}
\caption{Measured (a) and simulated (b) optical reflection measured through the quartz substrate for a 2x[Ti(10nm)Mg(20nm)] film covered
with 20 nm of Pd at different stages of loading. In the simulations two possible states are taken into account for the intermediate
loading: the ``loading from top'' and ``loading from bottom''. The dielectric functions of Ti, Mg, TiH$_2$, MgH$_2$ and Pd used in the
simulations are taken from Ref. \cite{Palik}, \cite{MachorroThinSolidFilms1995}, \cite{BorsaPRB2007}, \cite{IsidorssonPRB2003} and
\cite{Palik} respectively.}\label{spectra}
\end{center}
\end{figure}

\begin{figure}[htbp]
\begin{center}
\includegraphics[width=0.45\textwidth]{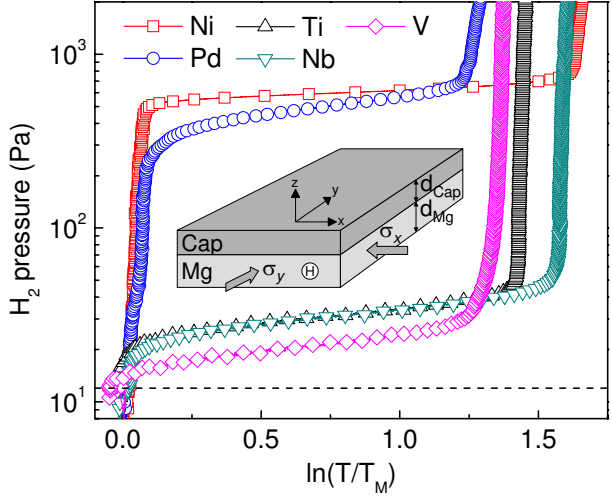}
\caption{(color online) Effect of cap-layer: PTI's measured at 333 K for Ti(10nm)Mg(20nm)X(10nm)Pd(10nm) samples deposited on glass with X
= Ni, Pd, Ti, Nb and V. The dashed line is the pressure at which coexistence of $\alpha$ and $\beta$ phases begins to appear upon hydrogen
absorption in bulk Mg \cite{Krozer1990}. In the inset: schematic visualization of the elastic model used to interpret the data.}\label{cap}
\end{center}
\end{figure}

\begin{figure}[htbp]
\begin{center}
\includegraphics[width=0.45\textwidth]{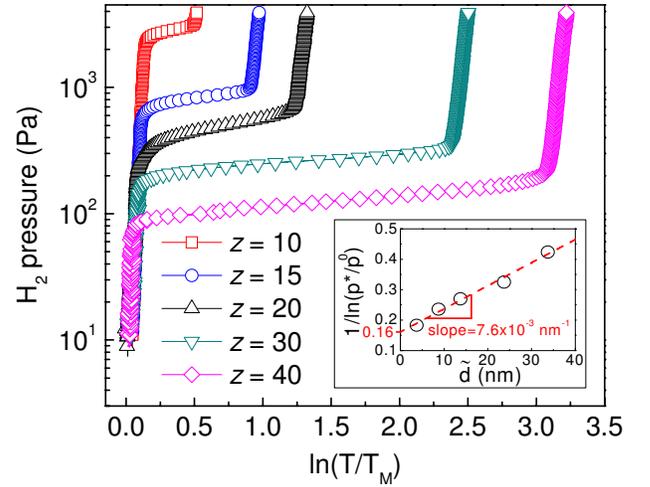}
\caption{(color online) Effect of Mg thickness: PTI's measured at 333 K for Ti(10nm)Mg($z$nm)Pd(40nm) samples deposited on glass with $z$ =
10, 15, 20, 30 and 40 nm. In the inset: Mg thickness dependence of the equilibrium pressure. The dashed line is a linear fit to the
points.}\label{thickness}
\end{center}
\end{figure}

In this letter we show that the thermodynamic stability of the Mg-H system can be drastically modified via elastic constraints. This is a
direct consequence of the long range H-H interaction in metals \cite{zabel1979PRL,Wagner1974}. In a metal with free surfaces two hydrogen
atoms feel an attractive interaction, while in a perfectly clamped sample the interaction becomes repulsive. Clamping can thus be used to
tune the equilibrium hydrogen pressure to the levels required for specific applications. To prove the validity of our approach we prepared
two series of capped Mg thin films: 1) to explore the role of chemical binding we deposited Mg thin films capped with different transition
metals: Ni, Pd, Ti, Nb and V; 2) in order to study the thickness dependence of the elastic effect we prepared a series of Pd-capped Mg
films of different thicknesses. The effect of clamping is strongly dependent on the chemical nature of the transition metal used as a
cap-layer and will be interpreted with a simple elastic model. By choosing the appropriate sample geometry we are able to increase the
hydrogen plateau pressure more than 200 times with respect to bulk Mg. Hydrogen absorption in our samples is measured by means of optical
spectroscopy and Hydrogenography \cite{Gremaud2007ADM}, a novel technique that allows to measure Pressure-Optical Transmission-Isotherms
(PTI's). In a PTI the amount of light transmitted by a thin film is measured as a function of increasing pressure at constant temperature:
when the metallic Mg films load with hydrogen the Mg-MgH$_2$ metal-insulator transition leads to an abrupt increase in the amount of
transmitted light. From the Beer-Lambert law, $\ln (T/T_{\mathrm{M}})\propto c_Hd$, the logarithm of the optical transmission, $T$,
normalized for the transmission of the film in its metallic state, $T_{\mathrm{M}}$, is proportional to the hydrogen concentration in the
film, $c_{\mathrm{H}}$ and to the film thickness, $d$. Measuring PTI's is therefore in all respects equivalent to measure standard
Pressure-Composition-Isotherms \cite{Gremaud2007APL}. The films used in this study are deposited in a UHV chamber ($\mathrm{p}<10^{-6}$ Pa)
by RF/DC magnetron sputtering on glass and quartz substrates. All films are covered with Pd to prevent oxidation and catalyze hydrogen
absorption. Optical reflection spectra are measured in a PerkinElmer Lambda 900 diffraction grating spectrometer, with energy range between
0.5 and 6.5 eV, during (de)hydrogenation at room temperature and H$_2$ pressures up to 10$^5$ Pa. PTI's are recorded in an optical cell
that allows to vary the hydrogen pressure from $10^{-1}$ to $10^6$ Pa at constant temperatures, between 313 and 573 K. Further details on
the hydrogenography experimental setup can be found in Ref. \cite{Gremaud2007ADM}. In Fig. \ref{bilayer} the PTI of a 2x[Ti(10nm)Mg(20nm)]
film covered with 20 nm of Pd is shown.
In the pressure and temperature ranges explored Pd and Ni do not load with hydrogen. Ti, Nb and V layers on the contrary load at the
beginning of the pressure ramp but, due to the metallic nature of their hydrides, produce very small optical effects compared to Mg. The
observed plateaus are therefore only due to the formation of MgH$_2$ upon hydrogen absorption. The isotherm shown in Fig. \ref{bilayer}
exhibits two clear and distinct plateaus of similar width. Since the width of the plateau is proportional to the thickness of the material
the two plateaus must originate from the two distinct Mg layers. Furthermore, to obtain such a well defined double-plateau we have to
assume that the two Mg layers are elastically disconnected \cite{PeislinAlefeld} and have different site energies for hydrogen absorption.
To prove these hypotheses we measure the optical reflection of the 2x[Ti(10nm)Mg(20nm)]Pd(20nm) sample deposited on quartz, while slowly
increasing the hydrogen pressure. Optical spectra are measured through the transparent quartz substrate at near-normal incidence. In Fig.
\ref{spectra}a the measured spectra are shown for three different steps of loading: 1) as-deposited metallic sample, 2) at intermediate
loading and 3) fully loaded sample. The calculated spectra shown in Fig. \ref{spectra}b are generated with SCOUT \cite{TheissSCOUT} by
taking into account the volume expansion occurring upon hydrogenation of Mg and Ti and including a rough interface between the quartz
substrate and the first Ti layer.
In the simulation the intermediate step is due to the complete loading of one of the two Mg layers while the other remains
metallic. Comparison with the experimental spectra shows unambiguously that the Mg layer sandwiched between two Ti layers absorbs hydrogen
at a lower pressure than the one capped with Pd. This is rather counterintuitive since the ``Ti-sandwiched Mg'' lies on the bottom of the
sample and hydrogen has to diffuse through the top Mg layer, which remains in the diluted $\alpha$ phase, in order to reach the bottom one.
For this mechanism to take place the two Mg layers have to differ with respect to their thermodynamic properties and they must be
elastically disconnected by the presence of a Ti layer in between. This remarkable behavior is the result of the following H loading
sequence. Already at very low H$_2$ pressures Ti forms TiH$_2$. The consequent 25\%{} volume expansion, together with the positive enthalpy
of solution of Mg and Ti, leads to disconnected interfaces between Mg and TiH$_2$ layers. The Ti-sandwiched Mg layer is then effectively
elastically disconnected from the surrounding. However, since Pd alloys with Mg and loads at higher pressures, this reasoning is not
applicable to the Ti-Mg-Pd block. When the top Mg layer expands upon H absorption it feels the Pd elastic constraint, resulting in a higher
hydride formation enthalpy and consequently in a higher plateau pressure. To demonstrate the validity of this interpretation we considered
two series of samples: a) Mg thin films capped with different transition metals, and b) Pd-capped Mg thin film with different thicknesses.
Fig. \ref{cap} shows the PTI's for 5 different samples with the following geometry: Ti(10nm)Mg(20nm)X(10nm)Pd(10nm) with X = Ni, Pd, Ti, Nb
and V.
Due to the presence of a Ti layer between Mg and the substrate we can assume that the only clamping effect is due to the top capping layer.
Clearly two behaviors are distinguishable: 1) Mg alloy forming elements such as Ni and Pd have a strong effect on the thermodynamic
properties of Mg, leading to a much higher plateau pressure than the one measured for Mg films of equal thickness capped with Ti, Nb and V;
2) the latter elements, which are immiscible with Mg, effectively behave like ``scissors'' and lead to ``quasi-free'' Mg layers with
elastically disconnected interfaces. Assuming that the origin of the destabilization observed is only elastic in nature, we can build a
simple model consisting of two layers - a Mg and a cap-layer - glued to each other as shown in the inset in Fig. \ref{cap}. Hydrogen atoms
absorbed by the Mg layer isotropically expand the lattice of the metallic host, while the Mg feels a compressive stress in the $x$ and $y$
directions due to the cap layer. In the $z$ direction there is no stress built up by the clamping but we have to take into account the
strain along the $z$ direction due to the Poisson's ratio, $\nu$. The two layers are in vacuum and no effect of the substrate
is taken into account, basically assuming that the bottom Ti layer is a perfect scissor. By taking into account only elastic constraints we
can calculate the effective volume expansion of the Mg layer upon hydrogen absorption. Furthermore, the enthalpy of hydride formation has a
simple volume dependence through the bulk modulus, $B$ \cite{GriessenFeenstra1985}: $\mathrm{d} \Delta H/\mathrm{d} \ln
V=-B_{\mathrm{Mg}}V_{\mathrm{H}}$, where $V_{\mathrm{H}}$ is the partial molar volume of hydrogen in Mg. We can then derive an expression
for the plateau pressure of the capped magnesium layer, Mg$^*$, with respect to free bulk magnesium, Mg$^0$:
\begin{equation}\label{all}
\ln\left(\frac{p^*}{p^0}\right)
=\frac{E_{\mathrm{Mg}}}{\tilde{E}}\left[1-\nu_{\mathrm{Mg}}+\left(1-\nu_{\mathrm{Cap}}\right)\frac{E_{\mathrm{Mg}}}{E_{\mathrm{Cap}}}\frac{d_{\mathrm{Mg}}}{d_{\mathrm{Cap}}}\right]^{-1}
\end{equation}
where $\tilde{E}=\frac{9}{4}RTV_{\mathrm{Mg}}/V_{\mathrm{H}}^2$, $V_{\mathrm{Mg}}$ is the molar volume of Mg and $E=3B(1-2\nu)$ is the
Young's modulus. The equation can also be rewritten as $[\ln(p^*/p^0)]^{-1}=I+Sd_{\mathrm{Mg}}$ with
\begin{equation}\label{IS}
I=\frac{\tilde{E}}{E_{\mathrm{Mg}}}(1-\nu_{\mathrm{Mg}}),\hspace{12pt} S=\frac{\tilde{E}}{E_{\mathrm{Cap}}}(1-\nu_{\mathrm{Cap}})\frac{1}{d_{\mathrm{Cap}}}
\end{equation}
According to the model the equilibrium pressure for the capped Mg has a straightforward dependence on the thickness of the Mg layer,
$d_{\mathrm{Mg}}$. Fig. \ref{thickness} shows the PTI's for 5 samples with different Mg thicknesses: Ti(10nm)Mg($z$nm)Pd(40nm), with $z$ =
10, 15, 20, 30 and 40 nm.
Two effects are clearly visible: a) the width of the plateau, $w$, is proportional to $d_{\mathrm{Mg}}$ as expected from the Beer-Lambert
law, b) the plateau pressure decreases with increasing Mg thickness as predicted by the model. A 10 nm thick Mg film has an equilibrium
hydrogen pressure of 2.8$\cdot10^{3}$ Pa at 333 K, more than 200 times higher than bulk Mg at the same temperature ($p^0_{333\
\mathrm{K}}=12$ Pa \cite{Krozer1990}). The plateau width is measured by dividing each isotherm into three regions: the solid solution
($\alpha$ phase) before the plateau, the coexistence region ($\alpha +\beta$ phases) and the fully hydrogenated region ($\beta$ phase)
after the plateau and fitting each part with a straight line: the width is then given by the distance, in units of $\ln
(T/T_{\mathrm{M}})$, between the intersection points. With the same geometrical construction we obtain the equilibrium pressure, as the
pressure in the middle of the plateau. Ideally one would expect $w\rightarrow 0$ for $d_{\mathrm{Mg}}\rightarrow0$, however, although the
plateau width is found to be proportional to the Mg thickness, a linear fit shows a relation of the type $w=A(d_{\mathrm{Mg}}-d_0)$, with
$d_0=6$ nm. The presence of an intercept indicates that the total amount of Mg undergoing a phase transition to MgH$_2$ is thiner
than expected and that we can assume for all samples that $\sim$6 nm of Mg are ``lost'' due to alloying with the top Pd layer. In the
comparison with the model we will therefore take into account an effective Mg thickness $\tilde d=d_{\mathrm{Mg}}-d_0$.  The formation of
an interfacial Mg-Pd alloy is directly responsible for the strong clamping effect observed and is consistent with the difference in width observed in the two plateaus of Fig. \ref{bilayer}, corresponding exactly to 6 nm. The inset in Fig. \ref{thickness} shows the
thickness dependence of the plateau pressure of Pd-capped Mg layers: plotting $[\ln(p^*/p^0)]^{-1}$ versus the effective Mg thickness we
find a linear dependence as predicted by the model, with intercept $I=0.16$ and slope $S=7.6\cdot 10^{-3}$ nm$^{-1}$. Substituting into
equations \ref{IS} the literature values for the structural parameters of Mg, Pd and H ($V_H=2.24$ cm$^3$/mol, $V_{\mathrm{Mg}}=13.97$
cm$^3$/mol, $\nu_{\mathrm{Mg}}=0.29$, $\nu_{\mathrm{Pd}}=0.39$, $E_{\mathrm{Mg}}=45.2$ GPa, $E_{\mathrm{Pd}}=126$ GPa) and taking $T=333$ K
and $d_{\mathrm{Pd}}=40$ nm, we obtain the theoretical values $I_\mathrm{th}=0.27$ and $S_\mathrm{th}=2.1\cdot10^{-3}$ nm$^{-1}$. The
agreement between experiment and theory is surprising when considering the strong approximations made in the model: no substrate is taken
into account, the only interaction is elastic, hence neglecting surface energy contributions, and the layers are assumed to behave as
perfect elastic bodies. The rather large $S$ value obtained from the fit can be explained considering the alloying effect taking place at the
Mg/Pd interface and assuming that only a fraction of $\sim$10 out of 40 nm of Pd has a significant clamping effect on the Mg underneath. The discrepancy in the $I$ values could reflect the fact that we consider
the Young's modulus of metallic Mg: taking into account MgH$_2$, which is stiffer than Mg, would lead to a closer agreement between
experiment and theory. The elastic model developed for a double-layer can be easily extended to 3D Mg nanoparticles embedded in a skin of
hard material. Assuming Mg particles of 10 nm of diameter and a skin of 2 nm thickness, the model predicts an increase in hydrogen
equilibrium pressure of more than 2000 times with respect to bulk Mg at the same temperature. The effect of clamping described in the
present work is not only important in the perspective of hydrogen storage applications, but it is crucial in understanding the
thermodynamics of all hydrogen-absorbing capped thin films, from switchable mirrors \cite{Huiberts1996} to electrode materials for
batteries \cite{Niessen2005}, hydrogen detectors \cite{SlamanSAB2007} multilayers and superlattices \cite{anderssonPRB1997}.

In conclusion we have shown the possibility to tailor the thermodynamics of a metal-hydrogen system by means of elastic constraints. Thin
films of Mg are used as a model metal-hydrogen system. Elements immiscible with Mg behave like scissors, while Mg-alloy-forming elements
exert a clamping effect that leads to huge increases in hydrogen equilibrium pressures. Clamping arises as a consequence of alloying
between Mg and the capping layer and its effects can be understood on the basis of a simple elastic model. The possibility to tune the
thermodynamics of a metal-hydrogen system by elastic means offers attractive new possibilities for compact hydrogen storage.

This work is supported by the Technologiestichting STW, the Nederlandse Organisatie voor Wetenschappelijk Onderzoek (NWO) through the
Sustainable Hydrogen Programme of Advanced Chemical Technologies for Sustainability (ACTS) and the Marie Curie Actions through the project
Complex Solid State Reactions for Energy Efficient Hydrogen Storage (COSY:RTN035366).

\end{document}